\documentclass[twocolumn,superscriptaddress,showpacs]{revtex4}
\usepackage{graphicx}

\begin{document}

\title{One-dimensional chirality: strong optical activity in epsilon-near-zero metamaterials}

\author{Carlo Rizza}
\affiliation{Dipartimento di Scienza e Alta Tecnologia, Universit\`a dell'Insubria, Via Valleggio 11, 22100 Como, Italy}
\affiliation{Consiglio Nazionale delle Ricerche, CNR-SPIN, Via Vetoio 10, 67100 L'Aquila, Italy}

\author{Andrea Di Falco}
\affiliation{School of Physics and Astronomy, University of St. Andrews, North Haugh, St. Andrews KY16 9SS,
United Kingdom}

\author{Michael Scalora}
\affiliation{Charles M. Bowden Research Center RDMR-WDS-WO, RDECOM, Redstone Arsenal, Alabama 35898-5000, USA}

\author{Alessandro Ciattoni}
\affiliation{Consiglio Nazionale delle Ricerche, CNR-SPIN, Via Vetoio 10, 67100 L'Aquila, Italy}

\begin{abstract}
We suggest that electromagnetic chirality, generally displayed by $3$D or $2$D complex chiral structures, can occur in
$1$D patterned composites whose components are achiral. This feature is highly unexpected in a $1$D system which is geometrically achiral since its mirror image can always be superposed onto it by a $180$ deg rotation. We analytically evaluate from first principles the bi-anisotropic response of multilayered metamaterials and we show that the chiral tensor is not vanishing if the system is geometrically \emph{one-dimensional chiral}, i.e. its mirror image can not be superposed onto it by using translations without resorting to rotations. As a signature of $1$D chirality, we show that $1$D chiral metamaterials support optical activity and we prove that this phenomenon undergoes a dramatic non-resonant enhancement in the epsilon-near-zero regime where the magneto-electric coupling can become dominant in the constitutive relations.
\end{abstract}


\maketitle

The term chirality is generally used to express the asymmetric property of a 3D object which can not be superimposed onto its mirror image by translations or rotations. 3D chirality is an important feature in many organic molecules (for example, 19 of the 20 common amino acids that form proteins are chiral) and its associated phenomena have enormous impact in several branches of science encompassing molecular biology, life science, optics, crystallography and particle physics. The concept of 2D chirality also exists and a planar object is said chiral if it cannot be superposed onto its mirror image unless it is lifted from the plane. Although the 3D chirality is widespread in nature, examples of 2D chiral structures are very few \cite{Flat_land}.

In the context of metamaterial science, artificial chiral structures, whose underlying constituents exhibit intrinsic chiral asymmetry, have been investigated theoretically and experimentally by several groups in 3D \cite{Jelinek_01,Gansel,Plum_01,Wang_01} and 2D configurations \cite{Papakostas,Fedotov,1c,Zhang_09,1b}. Chiral metamaterials have attracted a good deal of attention since they can yield strong chiral bi-anisotropic response (due to cross-coupling between the magnetic polarization and the electric one), giant optical activity, asymmetric transmission \cite{1}, repulsive Casimir force \cite{Casimir} and negative refractive index \cite{1d,1e,1f,1g}. It is worth noting that electromagnetic chirality can also be present in the case where underlying constituents are not intrinsically chiral. In this case, the magneto-electric coupling results from the geometric chirality of the whole structure and the effect is driven by the radiation wave vector contributing to the overall chiral asymmetry (extrinsic chiralilty). 2D extrinsic chiral metamaterials have been first proposed by E. Plum et al. \cite{Plum_PRL} and, in the considered media, authors have observed large optical activity which is indistinguishable from that occuring in media whose constituents are intrinsically chiral.

In this Letter, we show that 1D systems can support a peculiar reciprocal bi-anisotropic response in the long wavelength regime. We evaluate the chiral tensor of a class of multilayered metamaterials whose constituents are intrinsically achiral and whose 1D pattering is not characterized either by the standard 2D or 3D chiral asymmetry. We obtain the intriguing result that the chirality tensor does not vanish if the structure exhibits \textit{one-dimensional chirality}, i.e. its mirror image can not be superposed onto it by using translations without resorting to rotations.  Recently, a tunable bi-anisotropic response has been predicted in stratified media hosting graphene sheets \cite{2}. In the general situation that we are considering here, the effect of the 1D chiral asymmetry along the layers stacking direction is so pronounced that a medium with a three-layered unit cell, which is the simplest 1D chiral structure, can be characterized by a marked first order spatial dispersion which mimics the bi-anisotropic response of a uniaxial omega medium \cite{Capolino}. In addition, we analytically prove that 1D chiral metamaterials show optical activity (circular birefringence and circular dichroism) and this phenomenon is enhanced in the epsilon-near-zero regime to the point to be as strong as in 2D and 3D chiral metamaterials. Such non-resonant enhancement is a consequence of the general fact that close to permittivity crossing points the nonlocal and possible nonlinear polarization contributions can be comparable or even greater then the local linear part of the dielectric response. For example, a nonlinear epsilon-near-zero metamaterial can support efficient nonlinear processes \cite{PRA_sca,Argyro} and novel classes of solitons \cite{PRA_ciatt}. On the other hand, in linear epsilon-near-zero metamaterials, the medium response can be strongly affected by nonlocal response and it can be exploited to boost nonlocal phenomena such as the excitation of additional waves \cite{nonloc}. Here we suggest that nonlocal effects due to 1D chirality are enhanced by the epsilon-near-zero condition without any resonant mechanisms (i.e. with no cavity or plasmonic effects).

\begin{figure}
\includegraphics[width=0.5\textwidth]{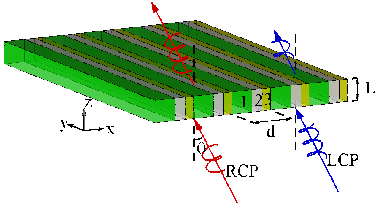}
\caption{(Color on-line) Sketch of metamaterial slab (with  $N=3$ layers) and
waves scattering geometry. The transmission amplitudes of right-handed (RCP,+)
and left-handed (LCP,-) circular polarized plane waves are not equal for $\theta \neq 0$ as
manifestation of 1D chirality.}
\end{figure}

Let us consider electromagnetic waves propagating in a 1D chiral metamaterial whose underlying structure has a unit cell obtained by stacking along the $x$-axis, N-layers of different media of thicknesses $d_j$ ($j=1,\dots,N$) (see Fig.1). The electromagnetic field amplitudes ${\bf{E}}$, ${\bf{H}}$, associated with monochromatic waves (the time dependence $e^{ - i \omega t}$ has been assumed where $\omega$ is the angular frequency) satisfy the Maxwell's equations $\nabla \times {\bf{E}}= i \omega \mu_0 {\bf{H}}$ and $\nabla \times {\bf{H}}= - i \omega {\bf{D}}$, where the constitutive relations are ${\bf{D}}=\epsilon_0 \epsilon {\bf{E}}$, ${\bf{H}}=(1/\mu_0)  {\bf{B}}$ and the relative dielectric permittivity $\epsilon$ is a periodic function of period $d = \sum \limits_{j = 1}^N {{d_j}}$. In order to obtain an effective medium description of the electromagnetic propagation valid in the regime where the ratio between the period d and the wavelength $\lambda$ is small, we exploit a standard multi-scale technique suitable for very general dielectric periodic profiles \cite{3,4}. Accordingly, we introduce the parameter $\eta  = d/\lambda \ll 1$ and the fast coordinate $X=x/\eta$ and, aimed at isolating the slowly and rapidly varying contributions, we consider the Fourier series of the dielectric permittivity and its inverse, i.e. ${\epsilon} = \left \langle {{\epsilon}} \right \rangle  + \sum \limits_{n \ne 0} {{a_n}{e^{in{k_0}X}}}$ and $\epsilon^{ - 1} = \left \langle {\epsilon^{ - 1}} \right \rangle + \sum \limits_{n \ne 0} {{b_n}{e^{in{k_0}X}}}$ (where $\left \langle f \right \rangle$ is the mean value of the function $f$ and ${k_0} = 2\pi / \lambda$). The basic Ansatz of our approach is ${\bf A} \left( {{\bf r},X} \right) = {{\bar {\bf A}}} \left( {\bf r} \right) +\delta { \bf A}\left( {{\bf r},X} \right) + \eta [{{\bar {\bf A}}}^{(1)} \left( {\bf r} \right) +\delta { \bf A}^{(1)} \left( {{\bf r},X} \right)]$, where $\bar{\bf A}$ and $\delta {\bf A}$ are the slowly (averaged) and rapidly varying part of each electromagnetic field ($\bf E$ or $\bf H$) whereas the superscript $(1)$ labels the first order corrections in $\eta$. Substituting the Fourier series of $\epsilon$ and $\epsilon^{-1}$ and the basic Ansatz into Maxwell equations, after separating the slowly and rapidly varying contributions and retaining the terms containing $\eta$ up to the first order, we obtain the equations describing the dynamics of the slowly varying parts of the electromagnetic field, namely $\nabla \times {\bar {\bf E}}= i \omega \mu_0 \bar {\bf H}$ and $\nabla \times {\bar{ \bf H}}= - i \omega {\bar {\bf D}}$, where we define
\begin{eqnarray} \label{3}
{\bar  D}_x&=&\epsilon_0  \left [ \epsilon_{||} {\bar  E}_x
+ \frac{ \kappa }{ k_0 }(\partial_y {\bar  E}_y +\partial_z {\bar  E}_z) \right ] ,\nonumber \\
{\bar  D}_y&=&\epsilon_0  \left( \epsilon_{\perp} {\bar  E}_y- \frac{ \kappa}{ k_0 } \partial_y {\bar  E}_x \right),\nonumber \\
{\bar  D}_z&=&\epsilon_0 \left( \epsilon_{\perp} {\bar E}_z-
\frac{\kappa}{k_0}\partial_z {\bar  E}_x \right),
\end{eqnarray}
$\bar{ \bf{H}}=(1/\mu_0)  \bar { \bf{B}}$, $\epsilon_{||}  = {\left\langle {\epsilon^{ - 1}} \right\rangle ^{ - 1}}$ and $\epsilon_{\perp} =
\left\langle {{\epsilon }} \right\rangle$ and we have introduced the chiral parameter
\begin{equation} \label{4}
\kappa  = i\eta \epsilon_{||} \sum\limits_{n \ne 0} {\frac{{{a_{ -n}}{b_n}}}{n}}.
\end{equation}
It is evident that in the limit $\eta \to 0$ the chiral parameter $\kappa$ vanishes and the multiscale approach considered in this paper reproduces the results of the well known standard effective medium theory.

The obtained expression for $\kappa$ of Eq.(\ref{4}) highlights the key role played by the fundamental concept of $1$D chirality. A structure is $1$D chiral if its mirror image can not be superposed onto it by using translations without resorting rotations.
From Eq.(\ref{4}) we have that the parameter $\kappa$ vanishes if the structure is 1D achiral. In fact, in this case, the permittivity profile after a reflection (say through the plane $X=0$) can be superposed to the original dielectric profile, i.e. it exists a translation $X'=X+ h$ such that $\epsilon(X) = \epsilon(-X + h)$. It is straightforward proving that the corresponding dielectric Fourier coefficients are such that $a_{-n} = \exp(-i n k_0 h) a_n$ and $b_{-n} = \exp(-i n k_0 h) b_n$ so that $a_{-n}b_{n}=a_{n}b_{-n}$ and the series of Eq.(\ref{4}) provides a vanishing $\kappa$. Therefore, the slowly varying and leading electromagnetic field can experience the effect of the novel terms proportional to its first order derivatives in the effective constitutive equations of Eq.(\ref{3}) only if the multilayer does exhibit 1D chiral asymmetry.

Note that Eqs.(\ref{3}) are of the kind $\bar{D}_i = \epsilon_0 \left( \epsilon_{ij} \bar{E}_j + \alpha_{ijn} \partial_n
\bar{E}_j \right)$ ($i=x,y,z$) describing media with a weakly spatial nonlocal dielectric response stemming from spatial dispersion \cite{5}. In the present case the components of the tensor $\alpha_{ijn}$ are all vanishing except $\alpha_{zxz} = - \alpha_{xzz} = - \kappa/k_0$,  $\alpha_{yxy} = - \alpha_{xyy} = -\kappa/k_0$ and this agrees with the fact that $\alpha$ satisfies the antisymmetric relation $\alpha_{ijn} = -\alpha_{jin}$ due to the Onsager symmetry
principle.

Exploiting the fact that the effective Maxwell's equations are invariant with respect to transformation $\bar{\bf D}'=\bar{\bf D}-\nabla \times {\bf Q}$, and $\bar{\bf H}'=\bar{\bf H}+ i\omega {\bf Q}$ (where $\bf Q$ is an arbitrary vector) \cite{44,44b}, after setting $ {\bf Q} =- \epsilon_0 \kappa/k_0 (\bar{E}_z {\hat {\bf e}}_y - \bar{E}_y {\hat {\bf e}}_z)$, we obtain the equivalent effective constitutive relations
\begin{eqnarray} \label{DH}
\bar{\textbf{D}} &=& \epsilon^{(eff)} \bar{\textbf{E}} - \frac{i}{c} \kappa^{(eff)T} \bar{\textbf{H}}, \nonumber \\
\bar{\textbf{B}} &=& \frac{i}{c} \kappa^{(eff)} \bar{\textbf{E}} + \mu^{(eff)} \bar{\textbf{H}},
\end{eqnarray}
where $\epsilon^{(eff)} = \textrm{diag} \left( \epsilon_{||},\epsilon_{\perp}+\kappa^2,\epsilon_{\perp}+\kappa^2 \right)$, $\mu^{(eff)}=\mu_0 I$ and
\begin{equation} \label{chiral-tens}
\kappa^{(eff)} = \left(
\begin{array}{ccc}
0 & 0 & 0 \\
0 & 0 & \kappa \\
0 & -\kappa & 0
\end{array}
\right)
\end{equation}
or, in other words, the considered metamaterial is a reciprocal bi-anisotropic metamaterial and, specifically, it mimics the response of an omega medium since the chirality tensor (reciprocal magneto-electric coupling) $\kappa^{(eff)}$ is purely antisymmetric \cite{Capolino,Trety}. Standard chiral materials have, by definition, chirality tensor with not-vanishing trace and therefore we again stress that, in this Letter, {\it one-dimensional chirality} is purely geometrical feature of structures whose mirror image can not be superposed onto them by using translations without resorting to rotations. It is worth noting that the structure of the chiral tensor of Eq.(\ref{chiral-tens}) of a one-dimensional chiral metamaterial can be directly obtained from symmetry considerations \cite{Arnaut}.

The stratified medium whose unit cell has two layers ($N=2$) has vanishing chiral tensor, $\kappa =0$, since it is always possible to superpose the mirror image of the structure with the original one through a suitable translation. Accordingly, as discussed in Ref.\cite{1h}, standard metal-dielectric stratified metamaterials show purely second order spatial dispersion. Therefore, the simplest stratified medium which can show not vanishing chiral tensor has three different layers in its unit cell with different permittivities $\epsilon_1,\epsilon_2,\epsilon_3$. The Fourier coefficient of the dielectric permittivity $\epsilon(X)$ are $a_n= \frac{i}{2\pi n} \left[ -(\epsilon_1-\epsilon_3) + (\epsilon_1-\epsilon_2) e^{-i 2\pi n f_1} + (\epsilon_2-\epsilon_3) e^{i 2 \pi n f_3} \right]$, where $n \neq 0$ and $f_i= d_i/d$ is the filling fraction of the $i$-th layer, and those of $1/\epsilon(X)$, $b_n$, have the same expression with the three permittivities $\epsilon_i$ replaced by their inverses $1/\epsilon_i$. Inserting these coefficients into Eq.(\ref{4}) and summing the resultant series using the relation $\sum_{n=1}^{\infty} \frac{\sin(2 \pi n f_i)}{n^3} = \frac{\pi^3}{3} \left(2 f^3_i -3 f^2_i + f_i \right)$ we obtain $\kappa = \pi \eta \epsilon _{||} \frac{(\epsilon_1 - \epsilon_2) (\epsilon_1 - \epsilon_3) (\epsilon_2 - \epsilon_3)} {\epsilon_1 \epsilon_2 \epsilon_3} f_1 f_2 f_3$.
\begin{figure}
\includegraphics[width=0.5\textwidth]{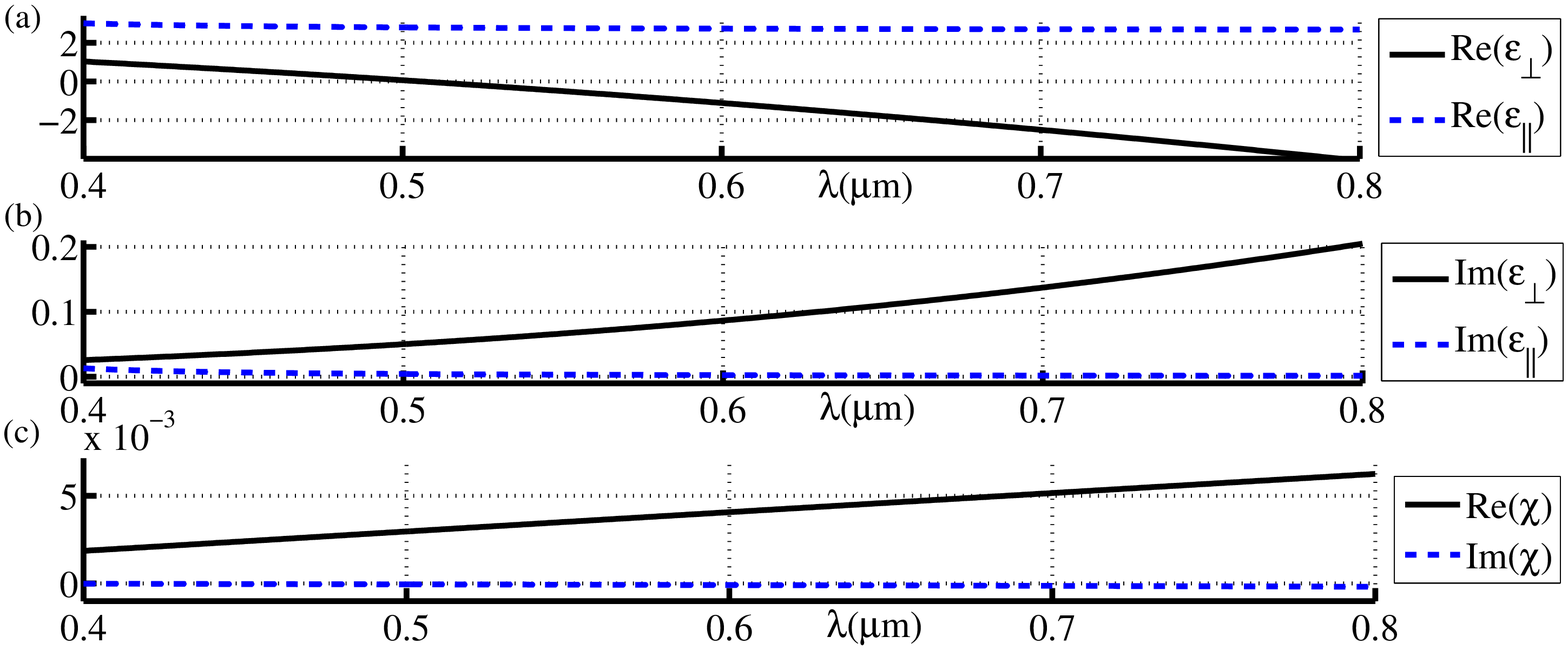}
\caption{Frequency dispersion of effective electromagnetic parameters.}
\end{figure}

As an example, we consider the electromagnetic effective response of a periodic three-layer structure with ${f_1} =0.18$, ${f_2} =0.33$, ${f_3}=(1-f_1-f_2)$, $d = 25$ nm, where we choose the slab layers of type 1, 2, 3 to be filled with Ag (1), SiO$_2$ (2), PMMA (3), respectively. The silver dielectric permittivity is given by the Drude model $\epsilon_1=\epsilon_b-\omega_p^2/(\omega^2+i\alpha \omega)$ (where $\epsilon_b=5.26$,
$\alpha=7.06 \cdot 10^{13}$ s$^{-1}$, $\omega_p=1.45 \cdot 10^{16}$ s$^{-1}$ \cite{JC}), whereas the frequency dispersion of SiO$_2$ ($\epsilon_2$) and PMMA ($\epsilon_3$) are described by empirical Sellmeier equations with different coefficients, respectively \cite{Malitson,info}. In Fig.2, we report the effective dielectric permittivities (panel (a) and (b)) and the chiral parameter $\kappa$ (panel (c)) as function of the wavelength $\lambda$. As shown in Fig. 2(a), the 1D material displays an effective dielectric permittivity with a zero-crossing point near $\lambda \simeq 0.51$ $\mu$m ($ \textrm{Re} (\epsilon_{\perp}) \simeq 0$) and it is characterized by a hyperbolic response ($Re(\epsilon_{\perp}) < 0$,
$Re(\epsilon_{||})> 0$) for $\lambda > 0.51$ $\mu$m.

The discussed electromagnetic bi-anisotropic response pertaining 1D chiral metamaterials is able to support specific optical activity effects such as circular birefringence and dichroism. Besides, the permittivity $\epsilon_\perp$ of the considered stratified structure, as above discussed, can have zero-crossing points (if both metal and dielectric layers are used) and therefore the epsilon-near-zero regime, i.e. $|\epsilon_\perp| \ll 1$, can occur in a spectral region around the crossing-points \cite{Rizzaaa,Enghets} where non-resonant enhancement of the chiral response is expected. In order to appreciate the effects of the $1$D chirality and their enhancement due to the epsilon-near-zero regime, we consider a slab of 1D chiral metamaterial of thickness $L$ illuminated by circularly polarized plane waves impinging from vacuum with incidence angle $\theta$ (see Fig.1). We here do not consider a substrate which, although important for practical implementations, does not provide an essential physical ingredient in the mechanism discussed here. For a given transversal wave vector $k_y$, there are two forward eigenwaves excited within the slab, i.e. the ordinary and extraordinary waves ${\bf E}^{(o)} = \frac{E^{(o)}}{k_0} \left(k_z^{(o)} \hat{\bf e}_y-k_y \hat{\bf e}_z \right) e^{i k_z^{(o)} z+i k_y y}$, ${\bf E}^{(e)} = E^{(e)}\left[\hat{\bf e}_x+i \frac{\kappa}{k_0 \epsilon_{\perp}}\left(k_y \hat{\bf e}_y+ k_z^{(e)}\hat{\bf e}_z\right)\right] e^{i k_z^{(e)} z+i k_y y}$, where $k_z^{(o)} = \sqrt{k_0^2 \epsilon_\perp-k_y^2}$ and $k_z^{(e)} = \sqrt{k_0^2\epsilon_{||} \left( 1 + \frac{\kappa^2}{\epsilon_\perp} \right)^{-1}-k_y^2}$. By imposing the continuity of tangential electric $\bar{\bf E}$ and magnetic $\bar{\bf H}'$ fields at the vacuum-metamaterial interfaces, one can obtain the complex transmission matrix $t$ relating the transmitted ${\bf E}^{(t)}$ and incident ${\bf E}^{(i)}$ circularly polarized fields, namely $E_l^{(t)}=t_{lm} E_m^{(i)}$, where indices $m$,$l$ correspond to polarization states of transmitted and incident wave, which can be either right (+) or left (-) circular polarizations. To quantify the optical activity of the considered metamaterial slab, we evaluate the difference in magnitude $\Delta=2 (|t_{--}|^2-|t_{++}|^2)/(|t_{--}|^2+|t_{++}|^2)$ and in phase $\delta \Phi=arg(t_{--})-arg(t_{++})$ of the two diagonal terms of the transmission matrix, since these two paramaters are associated to circular dichroism and circular birefringence, respectively. Using the effective permittivities and chiral parameter reported in Fig.2 for a slab of thickness $L = 0.4 \: \mu$m, in Fig.3 we plot the percentage difference $\Delta$ (panel (a)) and $\delta \Phi$ (panel(b)) as functions of the wavelength $\lambda$ for $\theta=0$ deg and $\theta=\pm 30$ deg. It is evident from Fig.3 that close to $\lambda \simeq 0.51$ $\mu$m, which corresponds to a crossing point of the permittivity $\epsilon_\perp$ (see panel (a) of Fig.2), both circular dichroism and circular birefringence show marked peaks.
\begin{figure}
\includegraphics[width=0.5\textwidth]{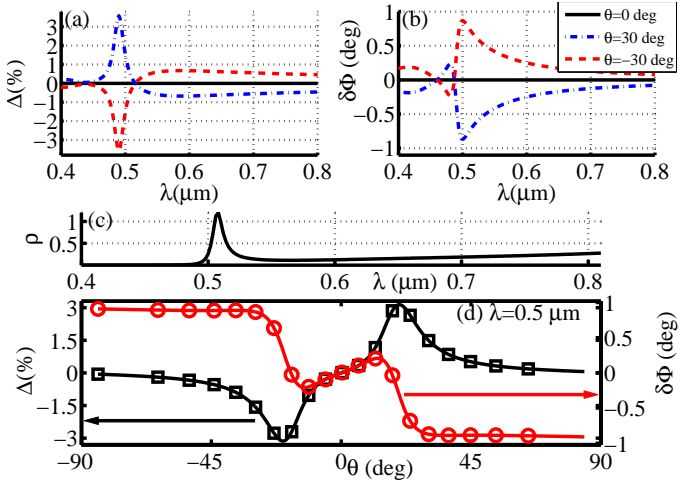}
\caption{Optical activity of a slab whose permittivities and chiral parameter are reported in Fig.2 and of thickness $L= 0.4 \: \mu m$. (a) Circular dichroism ($\Delta (\%)$) and (b) circular birefringence $\delta \Phi$ as functions of wavelength $\lambda$ for $\theta=0$ deg (solid line), $\theta=30$ deg (dash-dot line) and $\theta=-30$ deg (dash line). (c) Parameter $\rho$ measuring the magnitude of the magneto-electric coupling term relative to the local dielectric contribution from the $y$-component of the first of Eq.(\ref{DH}). (d) $\Delta (\%)$ (squares) and $\delta \Phi$ (circles) as function of $\theta$ for $\lambda=0.5$ $\mu$m.}
\end{figure}
\begin{figure}
\includegraphics[width=0.5\textwidth]{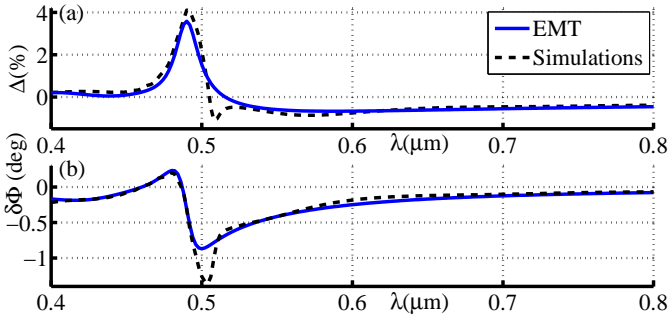}
\caption{Comparison between effective medium theory (EMT) (solid lines) and full-waves analysis (dashed lines) predictions about the optical activity of a metamaterial structure whose material parameters are those used for obtaining Fig.2 with $L=0.4$ $\mu$m and $\theta = 30$ deg.}
\end{figure}
From the first of Eqs.(\ref{DH}), it is worth noting that in the epsilon-near-zero regime $|\epsilon_\perp| \ll 1$ the magneto-electric coupling terms containing the chiral parameter can be comparable or greater then the local dielectric contributions and this triggers the non-resonant enhancement of the effects resulting from 1D chirality. This mechanism is proved in Fig.3(c) where we plot the parameter $\rho =  \left| \epsilon_0 \kappa^2  \bar{E}_y + i \frac{\kappa}{c} \bar{H}'_z \right| / \left| \epsilon_0 \epsilon_{\perp} \bar{E}_y \right|$, measuring the magnitude of the magneto-electric coupling term relative to the local dielectric contribution from the $y$-component of the first of Eq.(\ref{DH}) and evaluated at $z=L^{-}$ in the same situation of Fig.2(a-b), and $\rho$ shows a marked peak around $\lambda = 0.51$ $\mu$m reaching the maximum of $1.2$. The same mechanism can also be grasped from the expression of the extraordinary eigenwaves where it is evident that the first order parameter ruling the effect of 1D chirality on the electromagnetic field is $\kappa / \epsilon_\perp$ rather than $\kappa$ and this produces the enhancement of optical activity in the epsilon-near-zero regime. In Fig.3(d), we plot the percentage difference $\Delta$ and $\delta \Phi$ as functions of the angle $\theta$ for $\lambda = 0.5$ $\mu$m. Note that the circular dichroism and birefringence are completely absent for $\theta = 0$ deg (black solid line in Fig.3(a-b)) and they present opposite signs for opposite angles (dash-dot and dash lines in Fig.3(a-b)). These features are due to the fact that the $y$-component of the extraordinary wave (absent in local uniaxial materials) is proportional to  $k_y$. 1D chiral metamaterials share the same angular dependence observed in 2D extrinsic chiral metamaterials \cite{Plum_PRL} and they show optical activity analogous to one occurring in 2D and 3D chiral materials.

In order to check the validity of the above results we have performed 3D full-wave simulations \cite{Comsol} where two circularly polarized monochromatic plane waves are made to interact with the metamaterial structure as shown in Fig.1 and optical activity is accordingly retrieved. The considered structure has the same material parameters as those used to obtain Fig.2 and the scattering process has been numerically investigated for $L=0.4$ $\mu$m and $\theta = 30$ deg. In Fig.4 the comparison between the results of the effective medium theory (EMT) of Eqs.(\ref{DH}) and those of the numerical simulations are plotted and their good agreement is evident, together with the ENZ enhancement of the optical activity. The only slight discrepancies appear in a narrow spectral range centered at the ENZ wavelength $\lambda = 0.51$ $\mu$m since, in such spectral range, the effects of the (here neglected) higher order contributions to the dielectric response (e.g. containing second order electric field derivatives in Eqs.(\ref{3})) are boosted (as shown in Ref.\cite{nonloc}) as well as those due to first order contributions discussed in the present letter. In addition, such discrepancies can also arise from using an homogenization theory performed for an unbounded medium in the presence of a slab which has boundaries.

In conclusion, we have shown that a multilayered structure exhibiting 1D chiral asymmetry along the stacking direction shows a reciprocal bi-anisotropic electromagnetic response whose effects are comparable to those associated with 2D and 3D complex omega media. Here the bi-anisotropic response is supported by 1D metamaterials which are much easier fabricated and theoretically investigated than 2D and 3D chiral metamaterials. In addition, we have proposed a strategy to enhance the optical activity stemming from 1D chirality in metamaterials by exploiting the epsilon-near-zero regime where the magneto-electric coupling contribution to the medium dielectric displacement is even greater then the standard dielectric permittivity one. Our findings could be essential for conceiving ultra-efficient bi-anisotropic photonic devices in the optical frequency range.

A Ciattoni and C Rizza thank the U.S. Army International Technology Center Atlantic for financial support (Grant No. W911NF-14-1-0315). A Di Falco acknowledges support from EPSRC (EP/I004602/1 and EP/J004200/1).

\end{document}